\documentclass[11pt,a4paper]{article}
\usepackage{amsmath}
\usepackage{amsfonts}
\usepackage{amssymb}
\usepackage{color}
\usepackage{graphicx}
\usepackage{wasysym}
\usepackage{multicol}
\usepackage[utf8]{inputenc}
\usepackage{authblk}

\newtheorem{theorem}{Theorem}[section]

\newtheorem{corollary}[theorem]{Corollary}

\newtheorem{lemma}[theorem]{Lemma}
\newtheorem{example}[theorem]{Example}

\newtheorem{problem}[theorem]{Problem}
\def\qed{\hfill $\Box$\medskip}

\def\cS{{\cal S}}

\def\diag{{\rm diag}}

\title{Optimal Bounds on Functions of Quantum States \\
under Quantum Channels}

\author[1]{Chi-Kwong Li \thanks{ckli@math.wm.edu}}
\author[1,2]{Diane Christine Pelejo \thanks{dppelejo@email.wm.edu}}
\author[3]{Kuo-Zhong Wang \thanks{kzwang@math.nctu.edu.tw}}

\affil[1]{Department of Mathematics,  College of William and Mary}
\affil[2]{Department of Applied Science,College of William and Mary}
\affil[3]{Department of Mathematics, National Chiao Tung University}

\date{}
\begin{document}
\maketitle

\begin{abstract}
Let $(\sigma_1,\sigma_2) \mapsto D(\sigma_1, \sigma_2)$ be a scalar function
such as the trace distance,
the fidelity, and the relative entropy, etc.
For two given quantum states $\rho_1$ and $\rho_2$, we determine optimal bounds for
$D(\rho_1, \Phi(\rho_2))$ for  $\Phi \in \cS$
for different classes of functions $D(\cdot, \cdot)$,
where $\cS$ is the set of unitary quantum channels, the set of mixed unitary
channels, the set of unital quantum channels, and the set of all quantum channels. Furthermore, we determine states $\sigma=\Phi(\rho_2)$ that attain the optimal values and discuss the uniqueness of such states.
\end{abstract}



\section{Introduction}

In quantum sciences research,  one often compares
a pair of quantum states $\rho_1, \rho_2$ by considering some scalar functions
$D(\rho_1,\rho_2)$.
For instance, in quantum information and quantum control, one would like to
measure the  `distance' between a state $\rho_1$ and another state $\rho_2$ which
go through a quantum channel or a quantum operation $\Phi$.
Suppose quantum states are represented as $n\times n$ density matrices, i.e.,
positive semi-definite matrices with trace one. The following measures are often used
\cite{11,10}:
\begin{equation}
({\rm tr}|\rho_1-\rho_2|^{2})^{1/2}, \quad \frac{1}{2}{\rm tr} |\rho_1 - \rho_2|, \quad
 \sqrt{2}\sqrt{1-{\rm tr}|\sqrt{\rho_1}\sqrt{\rho_2}|},
\end{equation}
which are known as the Hilbert-Schmidt (HS) distance, the trace distance and the Bures distance, respectively. Here  $|\rho|$ is the positive semidefinite square root of $\rho^{\dagger}\rho$.
In particular, the Bures distance is a function of the fidelity \cite{nc}
\begin{equation}
F(\rho_1,\rho_2) = {\rm tr} |\sqrt{\rho_1}\sqrt{\rho_2}|.
\end{equation}
The purpose of this paper is to study the following.

\medskip\noindent
\begin{problem} \label{1.1}
Let $D$ be a scalar function on a pair of quantum states.
Suppose $\rho_1, \rho_2$ are two quantum states
and $\mathcal{S}$ is a set of quantum channels. Determine the optimal bounds for
$D(\rho_1, \Phi(\rho_2))$ for $\Phi \in \mathcal{S}$,
and also the states $\sigma = \Phi(\rho_2)$ attaining the optimal bounds.
\end{problem}
\medskip

These optimal bounds provide insight on the geometry of certain sets of quantum states \cite{mmpz,lpp} and play an important role in quantum state discrimination \cite{1,2,3}. Physically, if a quantum state $\rho_2$ goes through some quantum channel $\Phi$, one would like to know $D(\rho_1,\Phi(\rho_2))$ for another fixed quantum state $\rho_1$. If $\Phi$ is under our control, a solution to this problem can help us select $\Phi$ to attain the maximum or minimum value for $D(\rho_1,\Phi(\rho_2))$. On the other hand, if we only know that $\Phi$ lies in a certain class of quantum channels, then the solution will tell us the range of values where $D(\rho_1,\Phi(\rho_2))$ lies.

Denote by $\mathcal{M}_n, \mathcal{H}_n, \mathcal{D}_n$ the set of $n\times n$
complex matrices, the set of $n\times n$ Hermitian matrices, and the set of
$n\times n$ density matrices, respectively. Then quantum channels are
trace preserving completely positive map $\Phi: \mathcal{M}_n \rightarrow \mathcal{M}_n$
with the operator sum representation
\begin{equation}
\Phi(X) = \sum_{j=1}^r F_j X F_j^{\dagger} \quad \hbox{ for all } X \in \mathcal{M}_n,
\end{equation}
where $F_1, \dots, F_r \in M_n$ satisfy $\sum_{j=1}^rF_j^{\dagger}F_j = I_n$.
The map $\Phi$ is a unitary channel if $r = 1$ and $F_1$ is unitary; it is
a mixed unitary channel if every $F_j$ is a multiple of a unitary matrix;
it is unital if $\Phi(I_n) = I_n$.

\medskip
In the next two sections, we will obtain results for two general classes of functions
$D(\cdot,\cdot)$. The first type of functions will
cover the Hilbert-Schmidt (HS) distance and the trace distance.
The second type will cover
the fidelity, the Bures distance, and also the relative entropy defined by
\begin{equation}
S(\rho_1||\rho_2) = {\rm tr} \rho_1 (\log \rho_1 - \log \rho_2).
\end{equation}
For each class of functions, we will give the complete solution of Problem \ref{1.1} when
$\mathcal{S}$ is the set of unitary quantum channels, the set of mixed unitary channels and the set of unital quantum channels. These will be done in the next two sections. We also consider the set of all quantum channels and obtain a complete answer for the first class of functions, and partial results for the second class of functions. Some concluding remarks and future research directions will be mentioned in Section 4.

In our discussion, we will let $\{E_{11}, E_{12}, \dots, E_{nn}\}$ denote the standard
basis for $\mathcal{M}_n$.
By the following result \cite[Theorem 3.6]{li2},
the solutions of Problem \ref{1.1} are the same for the
set of mixed unitary channels and the set of
unital channels.

\begin{lemma} \label{2.3}
Let $\rho,\sigma\in \mathcal{D}_n$. The following are equivalent.
\begin{enumerate}
\item[{\rm 1.}] There exists a mixed unitary quantum channel $\Phi$ such that $\Phi(\rho)=\sigma$.
\item[{\rm 2.}] There exists a unital quantum channel $\Phi$ such that $\Phi(\rho)=\sigma$.
\item[{\rm 3.}] $\lambda(\sigma) \prec \lambda(\rho)$.
\item[{\rm 4.}] There are unitary $U_1, \dots, U_n \in M_n$ such that
$\sigma = \frac{1}{n} (U_1^{\dagger}\rho U_1 + \cdots + U_n^{\dagger}\rho U_n)$.

\end{enumerate}
\end{lemma}

\section{Schur-Convex Functions}

For $p \ge 1$, define the Schatten $p$-norm of a Hermitian matrix $A$ by
\begin{equation}
\|A\|_p = ({\rm tr}|A|^p)^{1/p}.
\end{equation}
It is not hard to see that if $A$ has eigenvalues $a_1, \dots, a_n$, then
\begin{equation}
\|A\|_p = \ell_p(a_1, \dots, a_n) = \big\{\sum_{j=1}^n |a_j|^p\big\}^{1/p}.
\end{equation}
The Hilbert Schmidt norm is $\|\cdot\|_2$
and, up to a multiple, the trace norm is $\|\cdot\|_1$.

In \cite[Theorem 4]{mmpz}, the authors observed that
\begin{equation}
\max_{U \hbox{ is unitary }} \|\rho_1 - U\rho_2U^{\dagger}\|_1 =
\|\Lambda^\downarrow(\rho_1) - \Lambda^\uparrow(\rho_2)\|_1,
\end{equation}
and
\begin{equation}
\min_{U \hbox{ is unitary }} \|\rho_1 - U\rho_2U^{\dagger}\|_1 =
\|\Lambda^\downarrow(\rho_1) - \Lambda^\downarrow(\rho_2)\|_1,
\end{equation}
where $\Lambda^\downarrow(X)$ $($respectively, $\Lambda^\uparrow(X))$ denotes the diagonal matrix
having the eigenvalues of $X$ as diagonal entries arranged in descending order (respectively,
ascending order).

Actually, the same result holds if one replaces $\|\cdot\|_1$ by any unitary similarity invariant (u.s.i.)
norm $\|\cdot\|$, i.e., a norm $\|\cdot\|$
satisfying $\|UXU^{\dagger}\| = \|X\|$ for all $X \in \mathcal{H}_n$ and unitary $U \in \mathcal{M}_n$. To describe the full generalization of the result, we need the notion of
majorization and Schur-convexity. One may see the excellent monograph
\cite{maj} for the background.
We give the basic definition in the following to facilitate our discussion.

Recall that for $x, y \in \mathbf{R}^n$, we say that $x$ is majorized by $y$, denoted by
$x \prec y$ if the sum of the $k$ largest entries of $x$ is not larger than that of $y$
for $j = 1, \dots, n-1$, and the sum of the entries of $x$ equals to that of $y$.
A function $f: \mathbf{R}^n \rightarrow \mathbf{R}$ is Schur-convex if $f(x) \le f(y)$
whenever $x \prec y$. It is strictly Schur-convex if $f(x) < f(y)$ whenever
$x \prec y$ and $x \ne y$.

Denote by $\lambda(X) = (\lambda_1(X), \dots, \lambda_n(X))$
the vector of eigenvalues of $X \in \mathcal{H}_n$ arranged in descending order. In particular, if $X\in \mathcal{D}_n$, then $\lambda(X)$ is a vector in the set
\begin{equation}\label{omega}
\Omega_n = \{(x_1, \dots, x_n): x_1 \ge \cdots \ge x_n \ge 0, \ x_1 + \cdots + x_n = 1\}.
\end{equation}
We have the following.
\begin{theorem} \label{2.1}
Suppose the function  $D: \mathcal{D}_n\times \mathcal{D}_n \rightarrow \mathbf{R}$
is defined by $D(\sigma_1,\sigma_2)= d(\lambda(\sigma_1-\sigma_2))$ for a Schur-convex
function $d: \mathbf{R}^n \rightarrow \mathbf{R}$. Then
\begin{equation}
\max_{U \hbox{ is unitary }} D(\rho_1,U\rho_2U^{\dagger}) =
D(\Lambda^\downarrow(\rho_1),\Lambda^\uparrow(\rho_2)),
\end{equation}
and
\begin{equation}
\min_{U \hbox{ is unitary }} D(\rho_1,U\rho_2U^{\dagger}) =
D(\Lambda^\downarrow(\rho_1),\Lambda^\downarrow(\rho_2)).
\end{equation}
The maximum is attained at $U\rho_2U^{\dagger}$ if there exists a unitary $V$ such that $V\rho_1V^{\dagger} = \Lambda^\downarrow(\rho_1)$
and $VU\rho_2 U^{\dagger}V^{\dagger} =\Lambda^\uparrow(\rho_2)$. The maximum is attained at $U\rho_2U^{\dagger}$ if there exists a unitary $V$ such that $VU\rho_2 U^{\dagger}V^{\dagger} = \Lambda^\downarrow(\rho_2)$.
The converses of the two preceding statements are also true if $d$ is strictly Schur-convex.
\end{theorem}

Theorem \ref{2.1} provides a complete solution to Problem \ref{1.1} for the
the set $\mathcal{S}$ of unitary channels if $D(\sigma_1,\sigma_2) = d(\lambda(\sigma_1-\sigma_2))$
for a Schur-convex function $d(\cdot)$.  In particular, it provides information about the state $\sigma=\Phi(\rho_2)$ that attains the maximum and minimum values. For example, take $\rho_1 = \diag(.55, .45, 0)$ and
$\rho_2 = \diag(.35, .33, .32)$, and let $\|\cdot\|$
be any u.s.i. norm. Since all u.s.i. norms are Schur-convex, then
$\|\rho_1 - \rho_2\|$ and $\|\rho_1 - \diag(.32, .33, .35)\|$
will yield the minimum and maximum values in the set
$\{\|\rho_1 - U\rho_2 U^\dag\|: U \hbox{ unitary}\}.$
Furthermore, if we choose a norm $||\cdot||$ that corresponds to a strictly Schur-convex function such as the Schatten $p$-norm for
$p \in (1, \infty)$, then
the lower bound and upper bound can only occur at
the matrices $\rho_2$ and
$\diag(.32,.33,.35)$, respectively.
On the other hand, for the Schatten 1-norm, i.e., the trace norm,
the minimum may occur at other matrices such as
$U\rho_2 U^\dag = \diag(.33, .35, .32)$ and
the maximum may occur at other matrices such as
$U\rho_2 U^\dag = \diag(.33, .32, .35)$.
Another situation where the optimal is attained by multiple states
may arise when $\rho_1$ has repeated eigenvalues. For example, if $\rho=\frac{1}{n}I_n$,
then for any $\Phi\in \mathcal{S}$, $\Phi(\rho_2)$ attains the maximum/minimum.

Next, we turn to Problem \ref{1.1} for the set $\mathcal{S}$ of mixed unitary channels and
unital channels.
By Lemma \ref{2.3}, and the results in \cite{li1}, we have the following solution of
Problem \ref{1.1} if $D(\sigma_1,\sigma_2) = d(\lambda(\sigma_1-\sigma_2))$
for a Schur-convex function $d(\cdot)$ and $\mathcal{S}$ is the set of mixed unitary
channels or the set of unital channels. Furthermore, as shown in
Lemma \ref{2.3}, we can always construct
the mixed unitary channel of the form
\begin{equation}
\sigma \mapsto  \frac{1}{n} (U_1\rho U_1^{\dagger} + \cdots + U_n\rho U_n^{\dagger})
\end{equation}
for some unitary $U_1, \dots, U_n\in M_n$.

\begin{theorem} \label{2.4}
Suppose the function  $D: \mathcal{D}_n\times \mathcal{D}_n \rightarrow \mathbf{R}$
is defined by $D(\sigma_1,\sigma_2)= d(\lambda(\sigma_1-\sigma_2))$ for a Schur-convex
function $d: \mathbf{R}^n \rightarrow \mathbf{R}$. Let $\mathcal{S}$ be the set of mixed unitary channels
or the set of unital channels acting on $\mathcal{M}_n$. Then
\begin{equation}
\max_{\Phi \in \mathcal{S}} D(\rho_1,\Phi(\rho_2)) =
D(\Lambda^\downarrow(\rho_1),\Lambda^\uparrow(\rho_2))
\end{equation}
and
\begin{equation}
\min_{\Phi \in \mathcal{S}} D(\rho_1,\Phi(\rho_2)) =
D\left(\Lambda^\downarrow(\rho_1),\sum_{j=1}^n d_j E_{jj}\right),
\end{equation}
where $(d_1, \dots, d_n)$ is determined by the following algorithm:

\medskip
Step 0. Set $(\Delta_1, \dots, \Delta_n) = \lambda(\rho_1) - \lambda(\rho_2)$.

\medskip
Step 1. If $\Delta_1 \ge \cdots \ge \Delta_n$, then set $(d_1, \dots, d_n)
= \lambda(\rho_1) - (\Delta_1, \dots, \Delta_n)$ and
stop.

\hskip .5in Else, go to Step 2.

\medskip
Step 2. Let $1 \le j < k \le \ell \le n$ be such that
\[\Delta_1\ge \cdots \ge \Delta_{j-1} > \Delta_j = \cdots = \Delta_{k-1} < \Delta_k
= \cdots = \Delta_\ell \ne \Delta_{\ell+1}.\]
\hskip .5in Replace each $\Delta_j, \dots, \Delta_\ell$
by $(\Delta_j+ \cdots + \Delta_\ell)/(\ell-j+1)$,
and go to Step 1.

\smallskip\noindent
The maximum
is attained at $\Phi\in \mathcal{S}$ if
there exists a unitary $V$ satisfying
 $V\rho_1V^{\dagger} = \Lambda^\downarrow(\rho_1)$ and
$V\Phi(\rho_2)V^{\dagger} =\Lambda^\uparrow(\rho_2)$.
The minimum  is attained at $\Phi\in \mathcal{S}$ if  there exists
a unitary $V$ satisfying $V\rho_1V^{\dagger} = \Lambda^\downarrow(\rho_1)$ and $V\Phi(\rho_2)V^{\dagger} = (\sum_{j=1}^n d_j E_{jj})$.
The converses of the above two statements also hold if $d$ is strictly
Schur-convex.
\end{theorem}

Here is an example illustrating the construction of the vectors
$(d_1, \dots, d_n)$ in the theorem.

\begin{example}
Let $\rho_1 = \frac{1}{10}{\rm diag}(4,3,3,0)$ and
$\rho_2 = \frac{1}{10}{\rm diag}(5, 2, 2, 1)$.

\medskip\noindent
Apply Step 0. Set $(\Delta_1, \dots, \Delta_4) =
\frac{1}{10}(4,3,3,0) -
\frac{1}{10}(5,2,2,1) =   \frac{1}{10}(-1,1,1,-1)$.

\medskip\noindent
Apply Step 2. Change $(\Delta_1, \dots, \Delta_4)$ to $\frac{1}{10}(1/3,1/3,1/3,-1)$.

\medskip\noindent
Apply Step 1. Set $(d_1, \dots, d_4) =  \frac{1}{10}(4,3,3,0) -
\frac{1}{10}(1/3,1/3,1/3,-1) = \frac{1}{30}(11,8,8,3)$.

\end{example}

Finally,  we consider the set $\mathcal{S}$ of all quantum channels. It is known that for any two quantum states, there is a quantum channel sending the first one to the second one.
We have the following.

\begin{theorem} \label{2.2}
Suppose the function  $D: \mathcal{D}_n\times \mathcal{D}_n \rightarrow \mathbf{R}$
is defined by $D(\sigma_1,\sigma_2)= d(\lambda(\sigma_1-\sigma_2))$ for a Schur-convex
function $d: \mathbf{R}^n \rightarrow \mathbf{R}$. Let $\mathcal{S}$ be the set of all quantum channels
acting on $\mathcal{M}_n$. Then
\begin{equation}
\max_{\Phi \in \mathcal{S}} D(\rho_1,\Phi(\rho_2)) =
D(\Lambda^\downarrow(\rho_1),E_{nn})
\quad \hbox{ and } \quad
\min_{\Phi \in \mathcal{S}} D(\rho_1,\Phi(\rho_2)) = D(\rho_1,\rho_1).
\end{equation}
The minimum is attained at $\Phi \in \cS$ if $\Phi(\rho_2) = \rho_1$.
The maximum is attained at $\Phi \in \cS$ if
there exists a unitary $V$ satisfying
$V\rho_1V^{\dagger} = \Lambda^\downarrow(\rho_1)$
and $V\Phi(\rho_2)V^{\dagger} = E_{nn}$.
If, in addition, $d$ is strictly Schur-convex, then the converses of the two preceding statements are also true.
\end{theorem}

\it Proof. \rm
The conclusion on the minimum is clear.
For the maximum, note that for any  $\sigma \in \mathcal{D}_n$,
\[\sum_{j=1}^k \lambda_j(\rho_1 - \sigma)
\le \sum_{j=1}^k \lambda_j(\rho_1) + \sum_{j=1}^k \lambda_j(-\sigma)
\le \sum_{j=1}^k \lambda_j(\rho_1) + \sum_{j=1}^k\lambda_j(-E_{nn})=\sum_{j=1}^k\lambda_j(\Lambda^\downarrow(\rho_1)-E_{nn})\]
for $j = 1, \dots, n-1$, and $\sum_{j=1}^n \lambda_j(\rho_1-\sigma) = 0$.
Because $d(\cdot)$ is Schur-convex, the result follows.
\qed

\section{Fidelity, relative entropy, and other functions}

In this section, we consider Problem \ref{1.1} for other functions
including the fidelity
\begin{equation}
F(\rho_1, \rho_2)= {\rm tr}\left(\sqrt{\sqrt{\rho_2} \rho_1\sqrt{\rho_2}}\right)
=\|\sqrt{\rho_1}\sqrt{\rho_2}\|_1 = {\rm tr}|\rho_1^{1/2} \rho_2^{1/2}|,
\end{equation}
and the relative entropy
\begin{equation}
S(\rho||\sigma)={\rm tr}(\rho(\log\rho-\log\sigma)) = {\rm tr}(\rho\log \rho) - {\rm tr}(\rho \log \sigma).
\end{equation}
In \cite[Theorem 2.1]{zf}, it was shown that  if $\mathcal{S}$ is the set of unitary channels, then
\begin{equation}
\max\limits_{\Phi\in \mathcal{S}} F(\rho_1,\Phi(\rho_2))=F(\Lambda^\downarrow(\rho_1),\Lambda^\downarrow(\rho_2))=\sum\limits_{j=1}^{n} \sqrt{\lambda_j(\rho_1)}\sqrt{\lambda_j(\rho_2)},
\end{equation}
and
\begin{equation}
\min\limits_{\Phi \in \mathcal{S}} F(\rho_1,\Phi(\rho_2))=F(\Lambda^\downarrow(\rho_1),
\Lambda^\uparrow(\rho_2))=\sum\limits_{j=1}^{n} \sqrt{\lambda_{j}(\rho_1)}\sqrt{\lambda_{n-j+1}(\rho_2)}.
\end{equation}
If $\mathcal{S}$ is the set of unital channels, it was also shown in \cite[Corollary 2.4]{lpp} that the above minimum is also valid, but
determining the maximum is an open problem.

In the following,  we consider different functions $f$ and $g$ on
quantum states and study upper bounds and lower bounds for a
function $D: \mathcal{D}_n \times \mathcal{D}_n \rightarrow \mathbf{R}$ of the form
\begin{equation}\label{Dfuntion}
D(\rho_1,\rho_2) = {\rm tr} f(\rho_1)g(\Phi(\rho_2)) \qquad \hbox{ and } \qquad
D(\rho_1,\rho_2) = {\rm tr} |f(\rho_1)g(\Phi(\rho_2))|
\end{equation}
with $\Phi \in \mathcal{S}$ for different sets $\mathcal{S}$ of quantum channels.
The results will cover a number of important functions in quantum information research,
and the techniques based on the theory of majorization can be further extended to other
functions.

To present our results, we need some more definitions
and results in majorization (see \cite{maj}) to present our general theorem.

A scalar function $f: [0,1] \rightarrow \mathbf{R}$ can be
extended to $f: \mathcal{D}_n \rightarrow \mathcal{H}_n$ such that
$f(\sigma) = U^{\dagger}{\rm diag}(f(\mu_1), \dots, f(\mu_n))U$ if $\sigma = U^{\dagger}{\rm diag}(\mu_1, \dots, \mu_n)U$,
where $U$ is unitary and $\mu_1 \ge \cdots \ge \mu_n \ge 0$.

For two vectors $x, y \in \mathbf{R}^n$,
$x$ is weakly majorized by $y$, denoted by $x \prec_w y$
if the sum of the $k$ largest entries of $x$ is not larger than that of $y$ for $k = 1, \dots, n$.
Furthermore, for $x, y \in \mathbf{R}^n$ with nonnegative entries,
$x$ is log majorized by $y$, denoted by $x \prec_{\log} y$
if the product of the entries of $x$ is the same as that of $y$,
and the product of the $k$ largest entries
of $x$ is not larger than that of $y$ for $k =1 , \dots, n-1$.
It is known that $x \prec_{\log} y$ then $x \prec_w y$.

We have the following.

\begin{theorem} \label{3.1}
Let $f,g:[0,1]\rightarrow \mathbf{R}$,  $\rho_1, \rho_2 \in \mathcal{D}_n$.
\begin{itemize}
\item[{\rm (a)}] If $f(\rho_1)$ and $g(\rho_2)$ have eigenvalues
$a_1 \ge \cdots \ge a_n$ and $b_1 \ge \cdots \ge b_n$, then
\begin{equation}
\min_{U \hbox{ unitary}} {\rm tr} (f(\rho_1)g(U^{\dagger}\rho_2U)) =
\sum_{j=1}^n a_jb_{n-j+1}, \
\max_{U \hbox{ unitary}} {\rm tr} (f(\rho_1)g(U^{\dagger}\rho_2U)) = \sum_{j=1}^n a_j b_j.
\end{equation}
The minimum is attained at a unitary $U$ if and only if there exists a unitary $V$ such that $V^{\dagger}f(\rho_1)V={\rm diag}(a_1, \dots, a_n)$ and
$V^{\dagger}g(U^{\dagger}\rho_2 U)V = g(V^{\dagger}U^{\dagger} \rho_2 UV) = {\rm diag}(b_n, \dots, b_1)$;
The maximum is attained at a unitary $U$ if and only if there exists a unitary $V$ such that $V^{\dagger}f(\rho_1)V={\rm diag}(a_1, \dots, a_n)$ and
$V^{\dagger}g(U^{\dagger}\rho_2 U)V
= g(V^{\dagger}U^{\dagger} \rho_2 UV) = {\rm diag}(b_1, \dots, b_n)$.

\item[{\rm (b)}]
 If $f(\rho_1)$ and $g(\rho_2)$ have singular values
$\alpha_1 \ge \cdots \ge \alpha_n$ and $\beta_1 \ge \cdots \ge \beta_n$, then
\begin{equation}
\min_{U \hbox{ unitary}} {\rm tr} |f(\rho_1)g(U^{\dagger}\rho_2U)| =
\sum_{j=1}^n \alpha_j\beta_{n-j+1},  \
\max_{U \hbox{ unitary}} {\rm tr} |f(\rho_1)g(U^{\dagger}\rho_2U)|
= \sum_{j=1}^n \alpha_j \beta_j.
\end{equation}
The minimum is attained at a unitary $U$ if and only if there exists a unitary $V$ such that
$|V^{\dagger}f(\rho_1)V|={\rm diag}(\alpha_1, \dots, \alpha_n)$ and
$|V^{\dagger}g(U^{\dagger}\rho_2 U)V| = |g(V^{\dagger}U^{\dagger} \rho_2 UV)| = {\rm diag}(\beta_n, \dots, \beta_1)$;
the maximum is attained at a unitary $U$ if and only if there exists a unitary $V$ such that $|V^{\dagger}f(\rho_1)V|={\rm diag}(\alpha_1, \dots, \alpha_n)$ and
$|V^{\dagger}g(U^{\dagger}\rho_2 U)V| = |g(V^{\dagger}U^{\dagger} \rho_2 UV)| = {\rm diag}(\beta_1, \dots, \beta_n)$.
\end{itemize}
\end{theorem}

\it Proof. \rm Let $f(\rho_1), g(\rho_2)$ have eigenvalues
$a_1 \ge \dots \ge a_n $ and $b_1 \ge \cdots \ge b_n $,
respectively.  For any unitary $V\in M_n$ satisfying $V^{\dagger}f(\rho_1)V = {\rm diag}(a_1, \dots, a_n)$, we have
\[{\rm tr}(f(\rho_1)g(U^{\dagger}\rho_2U))
= {\rm tr} ({\rm diag}(a_1, \dots, a_n) V^{\dagger}g(U^{\dagger}\rho_2U)V)
= {\rm tr} ({\rm diag}(a_1, \dots, a_n) g(V^{\dagger}U^{\dagger} \rho_2 UV)).\]
By \cite[II.9 Theorem H.1.g-h]{maj}, we have
\begin{equation}
\sum_{j=1}^n a_j b_{n-j+1} \le \sum_{j=1}^n a_j d_j \le \sum_{j=1}^n a_j b_j.
\end{equation}
Evidently, the bounds are attained if the unitary matrices
$U$ have the said properties.
Assertion (a) follows.

Next, suppose $f(\rho_1), g(\rho_2)$ have singular values
$\alpha_1 \ge \dots \ge \alpha_n \ge 0$ and $\beta_1 \ge \cdots \ge \beta_n \ge 0$,
respectively.  Suppose
$f(\rho_1)g(U^{\dagger}\rho_2U)$ has singular values $s_1, \dots, s_n$.
By \cite[II.9 Theorem H.1]{maj},
\begin{equation}
(\alpha_1 \beta_n, \dots, \alpha_n\beta_1) \prec_{\log} (s_1, \dots, s_n)
\prec_{\log} (\alpha_1\beta_1, \dots, \alpha_n\beta_n),
\end{equation}
and ${\rm tr}|f(\rho_1)g(U^{\dagger}\rho_2U)| = \sum_{j=1}^n s_j$ satisfies
\begin{equation}
\sum_{j=1}^n \alpha_j \beta_{n-j+1} \le \sum_{j=1}^n s_j \le \sum_{j=1}^n \alpha_j \beta_j.
\end{equation}
Suppose $V\in M_n$ is unitary such that
$V^{\dagger}f(\rho_1)V = {\rm diag}( \xi_1\alpha_1, \dots,  \xi_n\alpha_n)$ with $\xi_1,\ldots,\xi_n\in \{-1,1\}$.
One easily construct the unitary $U \in M_n$ so that
$g(U^{\dagger}\rho_2U)$ attaining the lower and upper bounds.
Evidently, only those unitary matrices having the said
properties will yield the optimal bounds.
Assertion (b) follows.
\qed

If $\mathcal{S}$ is the set of all unitary channels, then the lower bounds and upper bounds
in Theorem \ref{3.1} are attainable by ${\rm tr} f(\sigma_1)g(\Psi(\sigma_2))$ for some $\Psi \in \mathcal{S}$. There are no restrictions to the real valued functions
$f$ and $g$ in  Theorem \ref{3.1}. So, it can be applied to a wide variety of situations.
For example, if $f(x) = g(x) = \sqrt{x}$, we obtain the result for the fidelity function
$F(\sigma_1,\sigma_2) = {\rm tr}|f(\sigma_1)g(\sigma_2)|$ and conclude that
for any unitary $U \in M_n$,
\begin{equation}
\sum_{j=1}^n [\lambda_j(\rho_1) \lambda_{n-j+1}(\rho_2)]^{1/2} \leq F(\rho_1, U^{\dagger}\rho_2U)
\le \sum_{j=1}^n [\lambda_j(\rho_1) \lambda_{j}(\rho_2)]^{1/2}.
\end{equation}
If $f(x) = x$ and $g(x) = \log(x)$, then for any unitary $U \in M_n$,
\begin{equation}
\sum_{j=1}^n \lambda_j(\rho_1)\log\lambda_{n-j+1}(\rho_2)
\le {\rm tr} (\rho_1\log(U^{\dagger}\rho_2U))
\le \sum_{j=1}^n \lambda_j(\rho_1)\log\lambda_j(\rho_2).
\end{equation}
Here we use the convention that $0 \log 0 = 0$ and $a \log 0 = -\infty$ if $a \in (0,1]$.
Applying this result to $S(\sigma_1||\sigma_2) = {\rm tr} \sigma_1(\log \sigma_1-\log \sigma_2)$,
we have
\begin{equation}
\sum_{j=1}^n \lambda_j(\rho_1)\log(\lambda_j(\rho_1)/\lambda_{j}(\rho_2))
\le S(\rho_1||U^{\dagger}\rho_2U)
\le \sum_{j=1}^n \lambda_j(\rho_1)(\log(\lambda_j(\rho_1)/\lambda_{n-j+1}(\rho_2))
\end{equation}
for any unitary $U \in M_n$.

Next, we consider the set $\mathcal{S}$ of mixed unitary channels and the set of unital channels.
Given $\rho_1,\rho_2\in \mathcal{D}_n$, from Lemma \ref{2.3}, the following statements are true.

$(i)$ For any $\Phi \in \mathcal{S}$, we have
$\lambda(\Phi(\rho_2))\prec \lambda(\rho_2)$.

$(ii)$ If $f(\rho_1)$ has eigenvalues $a_1\ge \cdots\ge a_n\ge 0$, then for any $(x_1,\ldots,x_n)\prec \lambda (\rho_2)$,
there is $\Phi\in \mathcal{S}$ such that
\begin{equation}
{\rm tr} (f(\rho_1)g(\Phi(\rho_2))) = \sum_{j=1}^n a_j g(x_j), \quad \hbox{
and } \quad
{\rm tr} |f(\rho_1)g(\Phi(\rho_2))| = \sum_{j=1}^n |a_j g(x_j)|.
\end{equation}
Hence, we have the following.

\begin{theorem} \label{3.2}
Let $f,g: [0,1] \rightarrow \mathbf{R}$, $\rho_1, \rho_2 \in \mathcal{D}_n$, and $\Phi$ be a unital channel.
Suppose $f(\rho_1)$ have eigenvalues $a_1 \ge \cdots \ge a_n$, singular values $\alpha_1\ge \cdots\ge \alpha_n$, and
$\rho_2$ has eigenvalues $b_1\ge \cdots\ge b_n$.

\begin{itemize}
\item[{\rm (a)}]
The best lower upper and upper bounds of $\sum_{j=1}^n {\rm tr}(f(\rho_1)g(\Phi(\rho_2)))$ equal
\begin{equation}
\inf\big\{\sum_{j=1}^n a_j \lambda_{n-j+1}(g(\sigma)): \sigma\in \mathcal{D}_n, \
\lambda(\sigma) \prec (b_1, \dots, b_n)\big\},
\end{equation}
and
\begin{equation}
\sup\big\{\sum_{j=1}^n a_j \lambda_j(g(\sigma)):\sigma\in \mathcal{D}_n, \
\lambda(\sigma) \prec (b_1, \dots, b_n)\big\}, \quad \hbox{respectively}.
\end{equation}
Suppose the function $g(x)$ is increasing concave. Then the infimum
value $\sum_{j=1}^n a_j g(b_{n-j+1})$ is attainable, and
a unital channel $\Phi$ will attain the infimum value if and only if
there is a unitary $V$ satisfying
$V^{\dagger}f(\rho_1)V = {\rm diag}(a_1, \dots, a_n)$ and
$V^{\dagger}g(\Phi(\rho_2))V = {\rm diag}(g(b_n), \dots, g(b_1))$.
In particular, the infimum can be attained at a unitary channel.
\item[{\rm (b)}]
The best lower upper and upper bounds of $\sum_{j=1}^n {\rm tr}|f(\rho_1)g(\Phi(\rho_2))|$, respectively, are
\begin{equation}\
\inf\big\{\sum_{j=1}^n  \alpha_j \lambda_{n-j+1}(|g(\sigma)|):
\sigma\in \mathcal{D}_n, \
\lambda(\sigma) \prec (b_1, \dots, b_n)\big\},
\end{equation}
and
\begin{equation}
\sup\big\{\sum_{j=1}^n \alpha_j \lambda_j(|g(\sigma)|):
\sigma\in \mathcal{D}_n, \
\lambda(\sigma) \prec (b_1, \dots, b_n)\big\}.
\end{equation}
If the functions $f(x)$ and $g(x)$ have non-negative values on $[0,1]$, then the lower and
upper bounds are the same as those in {\rm (a)}. If in addition that
$g$ is increasing concave, then the infimum
value $\sum_{j=1}^n a_j g(b_{n-j+1})$ is attainable, and the infimum
will occur at $\Phi(\rho_2)$ that  satisfy the same conditions described
in {\rm (a)}.

\end{itemize}
\end{theorem}

\it Proof. \rm
(a) We may assume that $V^{\dagger}f(\rho_1)V = {\rm diag}(a_1, \dots, a_n)$.
For any
$\Phi \in \mathcal{S}$, we have
\begin{equation}
\sum_{j=1}^n a_j d_j={\rm tr} (f(\rho_1)g(\Phi(\rho_2))),
\end{equation}
where $(d_1, \dots, d_n)$ are the diagonal entries of $V^{\dagger}g(\Phi(\rho_2))V$.
Hence, $(d_1, \dots, d_n)$ is majorized by $\lambda (g(\Phi(\rho_2)))$, where
$\lambda (\Phi(\rho_2))$ are majorized by $(b_1, \dots, b_n)$.
 Similar to the proof of Theorem \ref{3.1},
\begin{equation}\sum_{j=1}^n a_j \lambda_{n-j+1}(g(\Phi(\rho_2)))\le \sum_{j=1}^n a_jd_j\le \sum_{j=1}^n a_j \lambda_{j}(g(\Phi(\rho_2)))
\end{equation}
Hence the forms of the best lower upper and upper bounds of $\sum_{j=1}^n {\rm tr}(f(\rho_1)g(\Phi(\rho_2)))$ holds.
If $g$ is increasing concave, we can apply (vi) of Table 2 in \cite[3.B.2]{maj} to the negative of the function
$\psi:(x_1,\ldots,x_n)\in \Omega_n\mapsto \sum_{j=1}^na_jg(x_{n-j+1})$
to show that $\psi$ is Schur-concave.
Thus the minimum occurs at $(x_1,\ldots,x_n)=(b_1,\ldots,b_n)$.

(b) Note that the singular values of $g(\Phi(\rho_2))$ are $\gamma_1 \ge \cdots \ge \gamma_n$,
which is a rearrangement of
$|g(x_1)|, \cdots |g(x_n)|$, where $x_1, \dots, x_n$ are the eigenvalues of $\Phi(\rho_2)$
satisfies  $(x_1, \dots, x_n) \prec (b_1, \dots, b_n)$.
Now, the eigenvalues of $|f(\rho_1) g(\Phi(\rho_2))|$ are the singular values of
$f(\rho_1)g(\Phi(\rho_2))$, which is log majorized by $(\alpha_1 \gamma_1, \dots, \alpha_n\gamma_n)$
and log majorizes $(\alpha_1\gamma_n, \dots ,\alpha_n \gamma_1)$. Thus,
\begin{equation}\sum_{j=1}^n \alpha_j \gamma_{n-j+1} \le {\rm tr} |f(\rho_1) g(\Phi(\rho_2))| \leq
\sum_{j=1}^n \alpha_j \gamma_{j}.
\end{equation}
If $f(x)$ has nonnegative values, then the eigenvalues of
$f(\rho_1)$ are the its singular values, and the same holds for $g(\Phi(\rho_2))$.
Thus, the results in (a) applies.
\qed

We can specialize the result to the function $f(x) = x$ and $g(x) = \log(x)$ to conclude that
$$\sum_{j=1}^n \lambda_j(\rho_1)\log \lambda_{n-j+1}(\rho_2))
\le {\rm tr} \rho_1 \log \Phi(\rho_2)$$
for any unital channel $\Phi$, and hence
\begin{equation}
S(\rho_1||\Phi(\rho_2)) =
{\rm tr} \rho_1 (\log \rho_1-\log \Phi(\rho_2)) \leq
\sum_{j=1}^n \lambda_j(\rho_1)\log(\lambda_j(\rho_1)/ \lambda_{n-j+1}(\rho_2)).
\end{equation}
For the Fidelity function
\begin{equation}
F(\rho_1, \Phi(\rho_2)) = {\rm tr} |\rho_1^{1/2} \Phi(\rho_2)^{1/2}|
\end{equation}
we can deduce the following result  in
\cite{lpp}
\begin{equation}
\min\limits_{\Phi\in \mathcal{S}} F(\rho,\Phi(\sigma))=F(\Lambda^\downarrow(\rho),\Lambda^\uparrow(\sigma))=\sum\limits_{i=1}^{n} \sqrt{\lambda_i(\rho)}\sqrt{\lambda_{n-i+1}(\sigma)}.
\end{equation}
It was noted in \cite{lpp} that the maximum value is not easy to determine.
As shown in Theorem \ref{3.2}, the upper bound of
$F(\rho_1, \Phi(\rho_2)) = {\rm tr}|\rho_1^{1/2}\Phi(\rho_2)^{1/2}|$ is the same as
the upper bound of
${\rm tr}(\rho_1^{1/2}\Phi(\rho_2)^{1/2})$, and one needs to determine
\begin{equation}
\sup\{ \sum_{j=1}^n\lambda_j(\rho_1)^{1/2}x_{j}^{1/2}: x_1 \ge \cdots \ge x_n\ge 0, \
(x_1, \dots, x_n) \prec \lambda(\rho_2)\}.
\end{equation}
By the continuity of the function $f(x) = g(x) = \sqrt{x}$ and the compactness of the
set
$\mathcal{R} = \{ (x_1,\ldots,x_n): x_1 \ge \cdots \ge x_n\ge 0, \
(x_1, \dots, x_n) \prec \lambda(\rho_2) \},$
we see that supremum is attainable. On the other hand, the determination of
the maximum depends heavily on $\lambda(\rho_1)$ and $\lambda(\rho_2)$.
For instance, if $\lambda(\rho_2) = (1/n,\dots, 1/n)$, then
$\mathcal{R}$ is a singleton and $F(\rho_1, \Phi(\rho_2)) = {\rm tr} |\rho^{1/2}|/\sqrt{n}$.
If $\rho_2 = {\rm diag}(1,0,\dots,0)$, then $\mathcal{R}$ contains all quantum states, and
$F(\rho_1,\Phi(\rho_2)) = 1$ if $\Phi(\rho_2) = \rho_1$. On the other hand, if
$\rho_1 = I_n/n$, then $I_n/n \in \mathcal{R}$ for any $\rho_2$ so that
$F(\rho_1,\Phi(\rho_2)) = 1$ for some unital channel $\Phi$.

In the following, we describe how to determine the
unital channel $\Phi$  that gives rise to
max $F(\rho_1,\Phi(\rho_2))$ for given $\rho_1,\rho_2 \in\mathcal{D}_n$.
The result actually covers a larger class of  functions.

\begin{theorem} \label{3.3}
Let $D: \mathcal{D}_n \times \mathcal{D}_n \rightarrow \mathbf{R}$ be defined as follows.
\begin{itemize}
\item[{\rm (a)}] $D(\sigma_1,\sigma_2) = {\rm tr}( f(\sigma_1) g(\sigma_2))$ or
$D(\sigma_1,\sigma_2) = {\rm tr}|f(\sigma_1) g(\sigma_2)|$, where
$f(x) = x^p$ and $g(x) = x^q$ with $p, q >0$ such that
$p+q = 1$, or
\item[{\rm (b)}] $D(\sigma_1,\sigma_2) = {\rm tr}( f(\sigma_1) g(\sigma_2))$
with $f(x) = x$ and $g(x) = \log x$.
\end{itemize}
Suppose $\mathcal{S}$ is the set of mixed unitary channels
or the set of unital channels acting on $\mathcal{M}_n$.
If $\rho_1,\rho_2\in \mathcal{D}_n$ have eigenvalues $a_1\ge \cdots \ge a_n\ge 0$ and
$b_1\ge \cdots\ge b_n\ge 0$, respectively, then
\begin{equation}
\max_{\Phi \in \mathcal{S}} D(\rho_1,\Phi(\rho_2)) =
\sum_{j=1}^nf(a_j)g(d_j),
\end{equation}
where $(d_1, \dots, d_n)$ is determined by the algorithm below.

\medskip\noindent
If $\Phi\in \mathcal{S}$
such that there exists a unitary $V$ satisfying $V^{\dagger}\rho_1V=\Lambda^{\downarrow}(\rho_1)$ and $V^{\dagger}\Phi(\rho_2)V = (\sum_{j=1}^n d_j E_{jj})$, then the upper bound is attained.

\medskip\noindent
{\bf Algorithm for determining $d_1 \ge \cdots \ge d_n$}:\\
Step 0. If $a_r>0$ and $a_{r+1}=\cdots=a_n=0$,
let $a=(a_1,\ldots,a_r)$ and $b=(b_1,\ldots,b_r)$, and set\\
\phantom{Step 0.} $(d_{r+1},\ldots, d_n) =(b_{r+1},\cdots, b_n)$.
$($Here, if $a_n > 0$, then $r = n$ and $(d_{r+1}, \dots, d_n)$ is vacuous.$)$ \\
Step 1. Let $k\in \{1 \dots, r\}$ be the largest integer  such that
\begin{equation}
\frac{1}{a_1 + \cdots + a_k}(a_1, \dots, a_k)
\prec \frac{1}{b_1 + \cdots + b_k}(b_1, \dots, b_k).
\end{equation}
Step 2. Set $(d_1, \dots, d_k) =\frac{b_1+\cdots+b_k}{a_1+\cdots+a_k} (a_1, \dots, a_k)$.
Stop if $k = r$. Otherwise, change $r$ to \\ \phantom{Step 2.} $r-k$, $a = (a_{k+1}, \dots, a_r), b = (b_{k+1}, \dots, b_r)$;
repeat Steps 1 and 2.

\end{theorem}

Note that in Step 0 of the algorithm above, we can alternatively choose $(d_{r+1},\ldots, d_n)=(b_{r+1},\ldots,b_n)S$ for any doubly stochastic matrix $S$. Also, in Step 1, $a_1+\cdots + a_k\neq 0$. This implies that $b_1+\cdots+b_k\neq 0$ because otherwise, the maximality of the choice for $k$ in the previous iteration  will be contradicted.

By Theorem \ref{3.3}, we see that
$S(\rho_1||\Phi(\rho_2)) \ge {\rm tr} (\lambda_j(\rho_1)\log(\lambda_j(\rho_j)/d_j))$,
where we use the usual convention that $0\log 0 = 0$ and $a \log 0 = -\infty$ if $a > 0$. The proof  of Theorem \ref{3.3}
is quite involved, and will be presented in Section 4. we illustrate the
results in Theorem \ref{3.3} and Theorem \ref{3.2} in the following example.

\begin{example}
Let $\rho_1 = \frac{1}{10} {\rm diag}(4,3,3,0)$ and $\rho_2 = \frac{1}{10}{\rm diag}(5, 2, 2, 1)$.

Apply Step 0. Set $d_4 = 0.1$, $a = (.4, .3, .3)$ and $b = (.5, .2, .2)$.

Apply
Step 1. Because $(0.4,0.3)/0.7 \prec (0.5,0.2)/0.7$,
$(0.4,0.3,0.3) \prec (0.5,0.2,0.2)/0.9$,

\hskip 1in we set $(d_1,d_2,d_3) = (0.36,0.27,0.27)$, and stop.

\medskip
Hence, $(d_1,d_2,d_3,d_4) = (0.36,0.27,0.27,0.1)$.
For the set $\mathcal{S}$ of unital channels,
\[
\min_{\Phi\in \mathcal{S}} F(\rho_1,\Phi(\rho_2)) = (\sqrt{4},\sqrt{3},\sqrt{3},0)(1,\sqrt{2},\sqrt{2},\sqrt{5})^t/10 = (2+2\sqrt{6})/10,\]
\[
\max_{\Phi\in \mathcal{S}} F(\rho_1,\Phi(\rho_2)) = (\sqrt{4},\sqrt{3},\sqrt{3},0)(\sqrt{3.6},\sqrt{2.7},\sqrt{2.7},1)^t/10 = 3/\sqrt{10}\]
\qquad and
\[
\min_{\Phi\in \mathcal{S}} S(\rho_1||\Phi(\rho_2))
= (4,3,3)(\log(10/9),\log(10/9),\log(10/9))^t/10,\]
\[\max_{\Phi\in \mathcal{S}} S(\rho_1||\Phi(\rho_2))
= (4,3,3)(\log 4,\log(3/2),\log(3/2))^t/10.\]
\end{example}

\medskip
Next we consider the set $\mathcal{S}$ of all quantum channels.
It is known that for any $\sigma_1, \sigma_2 \in \mathcal{D}_n$, there is a quantum channel $\Phi$
such that $\Phi(\sigma_1) = \sigma_2$.
Similar to Theorem \ref{3.2}, we have the following.

\begin{theorem} \label{3.5}
Suppose $\Omega_n$ is defined  as in (\ref{omega}),
$f,g: [0,1] \rightarrow \mathbf{R}$, and
$\rho_1, \rho_2 \in \mathcal{D}_n$ are such that
$f(\rho_1)$ have eigenvalues $a_1 \ge \cdots \ge a_n$ and singular
values $\alpha_1\ge \cdots\ge \alpha_n$.
Let $\Phi$ be a
quantum channel.
\begin{itemize}
\item[{\rm (a)}]
The best lower and upper bounds of $\sum_{j=1}^n {\rm tr}(f(\rho_1)g(\Phi(\rho_2)))$ equal
\begin{equation}
\inf\big\{\sum_{j=1}^n a_j \lambda_{n-j+1}(g(\sigma)):\lambda(\sigma)=
(x_1, \dots, x_n) \in \Omega_n \big\}
\end{equation}
and
\begin{equation}
\sup\big\{\sum_{j=1}^n a_j \lambda_j(g(\sigma)):\lambda(\sigma)=
(x_1, \dots, x_n) \in \Omega_n \big\}, \quad \hbox{respectively}.
\end{equation}
Suppose $g(x)$ is increasing concave, then the infimum value
equal to $g(0) \sum_{j=1}^{n-1} a_j  + a_n g(1)$ is attainable, and
$\Phi\in \cS$ attains the infimum if and only if there is a unitary $V$ such that
$V^{\dagger}f(\rho_1)V = {\rm diag}(a_1, \dots, a_n)$ and
$V^{\dagger}g(\Phi(\rho_2))V = {\rm diag}(g(0), \dots, g(0), g(1))$.
\item[{\rm (b)}]
The best lower and upper bounds of $\sum_{j=1}^n {\rm tr}|f(\rho_1)g(\Phi(\rho_2))|$ equal
\begin{equation}
\inf\big\{\sum_{j=1}^n \alpha_j \lambda_{n-j+1}(|g(\sigma)|):\lambda(\sigma)=
(x_1, \dots, x_n) \in \Omega_n \big\}
\end{equation}
and
\begin{equation}
\sup\big\{\sum_{j=1}^n \alpha_j \lambda_j(|g(\sigma)|):\lambda(\sigma)=
(x_1, \dots, x_n) \in \Omega_n \big\}, \quad \hbox{respectively}.
\end{equation}
If the functions $f(x)$ and $g(x)$ have non-negative values on $[0,1]$, then
the lower and upper
bounds are the same as that in {\rm (a)}. If in addition that
$g$ is increasing concave, then the infimum value
equals
${\rm tr} (f(\rho_1)g(\Phi(\rho_2))) =  g(0) \sum_{j=1}^{n-1} a_j  + a_n g(1)$
is attainable, and will occur at $\Phi(\rho_2)$
satisfying the same conditions as in {\rm (a)}.
\end{itemize}
\end{theorem}

In \cite{lpp}, it was proved that if $\mathcal{S}$ is the set of all quantum channels, then
\begin{equation}
 \max\limits_{\Phi\in\mathcal{S}} F(\rho_1,\Phi(\rho_2))=F(\rho_1,\rho_1)=1
\quad \hbox{ and } \quad
\min\limits_{\Phi\in \mathcal{S}} F(\rho_1,\Phi(\rho_2))=
\lambda_{\min}(\rho_1)^{\frac{1}{2}}.
\end{equation}
By Theorem \ref{3.5} and Lemma \ref{lemma} in the next section, we have the following.

\begin{corollary}\label{3.6}
Suppose $\mathcal{S}$ is the set of all quantum channels, and
 $\rho_1, \rho_2 \in \mathcal{D}_n$ have eigenvalues
$a_1 \ge \cdots \ge a_n \ge 0$ and $b_1 \ge \cdots \ge b_n \ge 0$, respectively.
The the following statements hold.
\begin{enumerate}
\item[{\rm (a)}] If $D(\sigma_1,\sigma_2) =
{\rm tr}(f(\sigma_1)g(\sigma_2))$ or $D(\sigma_1,\sigma_2) =
{\rm tr}|f(\sigma_1)g(\sigma_2)|$ with
$f(x)=x^p,g(x)=x^q$ such that $p,q>0$ and $p+q=1$, then
$$
\max_{\Phi\in \mathcal{S}} D(\rho_1,\Phi(\rho_2))=1\quad \hbox{ and } \quad
\min_{\Phi\in \mathcal{S}} D(\rho_1,\Phi(\rho_2))=f(a_n).
$$
\item[{\rm (b)}] For the relative entropy function,
\begin{equation}
\max_{\Phi\in \mathcal{S}}S(\rho_1||\Phi(\rho_2))=\infty \quad \hbox{ and } \quad \min_{\Phi\in \mathcal{S}}S(\rho_1||\Phi(\rho_2))=0.
\end{equation}
\end{enumerate}
\end{corollary}

\it Proof. \rm Similar to the proof of Theorem \ref{3.1}, we can focus on
\begin{equation}
 \sum_{j=1}^n f(a_j)g(z_j) \quad \hbox{ and } \quad
\sum_{j=1}^na_j\log a_j-\sum_{j=1}^n a_j\log z_j
\end{equation}
over the set $\Omega_n = \{(x_1, \dots, x_n): x_1 \ge \cdots \ge x_n\ge 0, \
x_1 + \cdots + x_n = 1\}$.

(a) \ The lower bound follows readily from Theorem \ref{3.2}. For the upper bound,
by Lemma \ref{lemma}(b), we have
\begin{equation}
\sum_{j=1}^nf(a_j)g(z_j)\le \sum_{j=1}^nf(a_j)g(a_j)=\sum_{j=1}^na_j^pa_j^q=1
\end{equation}
for all $(z_1,\ldots,z_n)\in \Omega_n$.

(b) \  Choose $(z_1,\ldots,z_n)=(0,\ldots,0,1)$. Since $a_1>0$, we have
\begin{equation}
\sum_{j=1}^na_j\log a_j-\sum_{j=1}^n a_j\log z_j=\infty.
\end{equation}
From Lemma \ref{lemma}(b),
$
\sum_{j=1}^na_j\log z_j\le \sum_{j=1}^na_j \log a_j
$
for all $(z_1,\ldots,z_n)\in \Omega_n$.
Hence
\begin{equation}
\min_{(z_1,\ldots,z_n)\in \Omega_n}(\sum_{j=1}^na_j\log a_j-\sum_{j=1}^n a_j\log z_j)=0.
\end{equation}
The result follows.
\qed

\section{Proof of Theorem \ref{3.3}}

To prove Theorem \ref{3.3}, we need some auxiliary results. We will focus
on the case when $a_1 \ge \cdots \ge a_n > 0$.

\begin{lemma}\label{lemma}
Suppose $f,g$ are defined as in Theorem {\rm \ref{3.3}}.
Given $p_1,\ldots,p_\eta,t_1\in [0,1]$ such that $p_1+\cdots+p_{\eta}> 0$, let
\begin{equation}
F_{p_1,\ldots,p_\eta,t_1}(x_1,\ldots,x_{\eta-1})=f(p_1)g(x_1)+\cdots+f(p_{\eta-1})g(x_{\eta-1})+f(p_\eta)g(t_1-x_1-\cdots-x_{\eta-1})
\end{equation}
for $ x_1,\ldots,x_{\eta-1}\geq 0$ and $x_1+\cdots+x_{\eta-1}\le t_1$.
Then the following statements are true:
\begin{enumerate}
\item[{\rm (a)}] $F_{p_1,p_2,t_1}(x_1)$ is concave for $x_1\in [0,t_1]$;
\item[{\rm (b)}] 
 $F_{p_1,\ldots,p_\eta,t_1}(x_1,\ldots,x_{\eta-1})<F_{p_1,\ldots,p_\eta,t_1}(\alpha p_1,\ldots,\alpha p_{\eta-1})$
     for all $(x_1,\ldots,x_{\eta-1})\neq \alpha (p_1,\ldots,p_{\eta-1})$, where $\alpha=\frac{t_1}{p_1+\cdots+p_\eta}$.
\end{enumerate}

\end{lemma}

\it Proof. \rm
For $\eta=2$, we have $F'_{p_1,p_2,t_1}(\frac{p_1t_1}{p_1+p_2})=0$ and $F^{''}_{p_1,p_2,t_1}(x_1)<0$ for all $x_1\in (0,t_1)$.
Hence, $(a)$ holds and in the case $\eta=2$, $(b)$ is true.
Assume that $\eta=k>2$.
$F_{p_1,\ldots,p_k,t_1}$ is continuous in $\Gamma_k\equiv \{(x_1,\ldots,x_{k-1}):0\le x_i\le t_1,x_1+\cdots+x_{k-1}\le t_1\}$.
Since $\Gamma_k$ is compact, there exists $(\hat{x}_1,\ldots,\hat{x}_{k-1})\in \Gamma_k$ such that
\begin{equation}
F_{p_1,\ldots,p_k,t_1}(\hat{x}_1,\ldots,\hat{x}_{k-1})=\max F_{p_1,\ldots,p_k,t_1}(\Gamma_k).
\end{equation}
From the case $\eta=2$, we get $\hat{x}_j=\frac{\hat{x}_j+\hat{x}_i}{p_j+p_i}p_j$ for all $i,j$ and $i\neq j$.
This implies that $\hat{x}_j=\alpha p_j$ for all $j$.
Since $\hat{x}_1+\cdots+\hat{x}_k=t_1$, we obtain $\alpha=\frac{t_1}{p_1+\cdots+p_k}$ and then $(b)$ holds.
\qed

\begin{theorem}\label{concave}
Let $f,g$ be defined as in Theorem {\rm \ref{3.3}}.
Suppose $a=(a_1,\ldots,a_n), b=(b_1,\ldots,b_n), x =(x_1,\ldots,x_n)$
are nonnegative decreasing sequences with $a_n > 0$ and that
$x\prec b$ satisfies
\begin{equation}
\sum_{j=1}^nf(a_j)g(x_j)\equiv f(a)g(x)\ge f(a)g(y)\quad \mbox{ for all }y\prec b.
\end{equation}
Then the following statements hold.
\begin{enumerate}
\item[{\rm (a)}] There exist $n_0=0<1\le n_1<n_2<\cdots<n_k=n$ such that for $0\le i<k$,
\[
\sum_{j=n_i+1}^{n_{i+1}}x_j=\sum_{j=n_i+1}^{n_{i+1}}b_j
\quad
\hbox{ and }
\quad (x_{n_i+1},\ldots,x_{n_{i+1}})=\alpha_i(a_{n_i+1},\ldots,a_{n_{i+1}}),\]
where $\alpha_i=\frac{b_{n_i+1}+\cdots+b_{n_{i+1}}}{a_{n_i+1}+\cdots+a_{n_{i+1}}}$.
\item[{\rm (b)}] The values $n_1, \dots, n_k$ in {\rm (a)}
can be determined as follows:
\begin{equation}
n_1=\max\{r:\alpha (a_1,\ldots,a_r)\prec (b_1,\ldots,b_r)\},
\end{equation}
and
\begin{equation}
n_j=\max\{r:\alpha (a_{n_{j-1}+1},\ldots,a_{r})\prec (b_{n_{j-1}+1},\ldots,b_{r})\} \quad
\hbox{ for } 1<j\le k.
\end{equation}
\end{enumerate}
\end{theorem}

\it Proof. \rm 
(a) Let $n_1=\max\{k: (x_1,\ldots,x_k)\prec \alpha_0(a_1,\ldots,a_k) \mbox{ for some }\alpha_0\}$.
If $n_1=n$, then by Corollary \ref{3.6}, the proof is done. 

Suppose that $n_1<n$.
Then $(x_1,\ldots,x_{n_1})=\alpha_0(a_1,\ldots,a_{n_1})$ and $x_{n_1+1}\neq \alpha_0 a_{n_1+1}$.
We claim that $\sum_{j=1}^{n_1}x_j=\sum_{j=1}^{n_1}b_j$.
Suppose that $\sum_{j=1}^{n_1}x_j<\sum_{j=1}^{n_1}b_j$.
Let $\beta=\frac{x_{n_1}+x_{n_1+1}}{a_{n_1}+a_{n_1+1}}$.
If $x_{n_1}=\beta a_{n_1}$, then $x_{n_1+1}=\beta a_{n_1+1}$.
Since $x_{n_1+1}\neq \alpha_0a_{n_1+1}$ and $x_{n_1}=\alpha_0a_{n_1}$, $\beta\neq \alpha_0$.
Thus, $\beta<\alpha_0$ or $\beta>\alpha_0$.

{\bf Case 1.}  $\beta<\alpha_0$.
Let $\hat{x}=(x_1,\ldots,x_{n_1-1},\beta a_{n_1},\beta a_{n_1+1},x_{n_1+2},\ldots,x_n)$.
We have $\beta a_{n_1}<\alpha_0 a_{n_1}=x_{n_1}$ and $\beta a_{n_1+1}=x_{n_1}+x_{n_1+1}-\beta a_{n_1}>x_{n_1+1}$.
Hence $\hat{x}$ is decreasing and $\hat{x}\prec b$.
On the other hand,
 \begin{eqnarray*}
&&f(a)g(\hat{x})-f(a)g(x)\\
 &=& f(a_{n_1})g(\beta a_{n_1})+f(a_{n_1+1})g(\beta a_{n_1+1})-(f(a_{n_1})g(x_{n_1})+f(a_{n_1+1})g(x_{n_1+1}))\\
 &=&F_{a_{n_1},a_{n_1+1},x_{n_1}+x_{n_1+1}}\left(a_{n_1}\frac{x_{n_1}+x_{n_1+1}}{a_{n_1}+a_{n_1+1}}\right)-F_{a_{n_1},a_{n_1+1},x_{n_1}+x_{n_1+1}}(x_{n_1})\\
 &>&0 \qquad (\mbox{by Lemma \ref{lemma}}   (b)).
 \end{eqnarray*}
This is a contradiction.

{\bf Case 2.} $\beta>\alpha_0$.
There exist $m_1\le n_1<m_2$ such that
\begin{equation}
x_{m_1-1}>x_{m_1}=\cdots=x_{n_1}\ge x_{n_1+1}=\cdots=x_{m_2}>x_{m_2+1}.
\end{equation}
We will show that $\sum_{j=1}^rx_j<\sum_{j=1}^rb_j$ for $m_1\le r<m_2$.

\medskip
\noindent
{\bf Assertion 1.}  $\sum_{j=1}^rx_j<\sum_{j=1}^rb_j$ for $n_1+1\le r<m_2$.

If not, then $\sum_{j=1}^{r_0}x_j=\sum_{j=1}^{r_0}b_j$ for some $n_1+1\le r_0<m_2$.
Because $\sum_{j=1}^{r_0+1}x_j\le \sum_{j=1}^{r_0+1}b_j$, we see that $x_{r_0+1}\le b_{r_0+1}$.
Since $\sum_{j=1}^{n_1}x_j<\sum_{j=1}^{n_1}b_j$, we may assume $\sum_{j=1}^rx_j<\sum_{j=1}^rb_j$
for $n_1\le r<r_0$.
We get $x_{r_0}>b_{r_0}\ge b_{r_0+1}$.
But $x_{r_0}=x_{r_0+1}\le b_{r_0+1}$.
This is a contradiction. Thus,
\begin{equation}
\sum_{j=1}^rx_j<\sum_{j=1}^rb_j \qquad \hbox{ for } n_1+1\le r<m_2.
\end{equation}
\medskip
\noindent
{\bf Assertion 2.} $\sum_{j=1}^rx_j<\sum_{j=1}^rb_j$ for $m_1\le r<n_1$.

If not, then $\sum_{j=1}^{r_1}x_j=\sum_{j=1}^{r_1}b_j$ for some $m_1\le r_1<n_1$.
Then $x_{r_1}\ge b_{r_1}$.
Since $x_{r_1}=\cdots=x_{n_1}$ and $b_{r_1}\ge \cdots\ge b_{n_1}$, we have $\sum_{j=1}^rx_j\ge \sum_{j=1}^rb_j$ for $r_1\le r\le n_1$.
This is impossible since $\sum_{j=1}^{n_1}x_j<\sum_{j=1}^{n_1}b_j$.
Hence $\sum_{j=1}^rx_j<\sum_{j=1}^rb_j$ for $m_1\le r<n_1$.

By the above argument, $\sum_{j=1}^rx_j<\sum_{j=1}^rb_j$ for $m_1\le r<m_2$.
Now, let
\begin{equation}
\hat{x}=(x_1,\ldots,x_{m_1-1},x_{m_1}+\delta,x_{m_1+1},\ldots,x_{m_2-1},x_{m_2}-\delta,x_{m_2+1},\ldots,x_n).
\end{equation}
For sufficiently small $\delta >0$, $\hat{x}$ is decreasing and $\hat{x}\prec b$.
In fact,
\begin{equation}
\alpha_0<\beta=\frac{x_{n_1}+x_{n_1+1}}{a_{n_1}+a_{n_1+1}}=\frac{x_{m_1}+x_{m_2}}{a_{n_1}+a_{n_1+1}}=\frac{x_{m_1}+x_{m_2}}{a_{m_1}+a_{n_1+1}}\le \frac{x_{m_1}+x_{m_2}}{a_{m_1}+a_{m_2}}.
\end{equation}
The third equality holds because $\alpha_0a_{m_1}=x_{m_1}=x_{n_1}=\alpha_0a_{n_1}$.
Hence
\begin{equation}
a_{m_1}\frac{x_{m_1}+x_{m_2}}{a_{m_1}+a_{m_2}}>\alpha_0 a_{m_1}=x_{m_1}.
\end{equation}
Then for sufficiently small $\delta>0$,
 \begin{eqnarray}
&&f(a)g(\hat{x})-f(a)g(x)\nonumber\\
 &=& f(a_{m_1})g(x_{m_1}+\delta)+f(a_{m_2})g(x_{m_2}-\delta)-(f(a_{m_1})g(x_{m_1})+f(a_{m_2})g(x_{m_2}))\nonumber\\
 &=&F_{a_{m_1},a_{m_2},x_{m_1}+x_{m_2}}(x_{m_1}+\delta)-F_{a_{m_1},a_{m_2},x_{m_1}+x_{m_2}}(x_{m_1})\nonumber\\
 &>&0 \qquad (\mbox{by Lemma \ref{lemma}} ).
 \end{eqnarray}
This is a contradiction and then $\sum_{j=1}^{n_1}x_j=\sum_{j=1}^{n_1}b_j$.
Let
\begin{equation}
n_2=\max\{k:(x_{n_1+1},\ldots,x_k)=\alpha (a_{n_1+1},\ldots,a_{k})\mbox{ for some }\alpha\}.
\end{equation}
From the above proof, we also have $\sum_{j=n_1+1}^{n_2}x_j=\sum_{j=n_1+1}^{n_2}b_j$.
By induction, we get the desired conclusion.

(b) Suppose $n_1<\eta\equiv \max\{r:\alpha (a_1,\ldots,a_r)\prec (b_1,\ldots,b_r)\}$.
We have
\begin{equation}
\sum_{j=1}^{n_1}x_j=\sum_{j=1}^{n_1}\alpha_0a_j=\sum_{j=1}^{n_1}b_j<\sum_{j=1}^{\eta}b_j
=\sum_{j=1}^{\eta}\alpha'a_j
\end{equation}
for some $\alpha'$.
Let $1< r<k$ with $n_{r-1}<\eta\le n_r$.
Then
\begin{equation}
\sum_{j=1}^{\eta}\alpha'a_j=\sum_{j=1}^\eta b_j\ge \sum_{j=1}^\eta x_j=\sum_{j=1}^{n_{r-1}}b_j+\sum_{j=n_{r-1}+1}^\eta x_j.
\end{equation}
There is $0<\alpha^{''}\le \alpha'$ such that $\sum_{j=1}^\eta \alpha^{''}a_j=\sum_{j=1}^\eta x_j$.
Then $\sum_{j=1}^p\alpha^{''}a_j\le \sum_{j=1}^pb_j$ for $1\le p\le \eta$.
We have $\sum_{j=1}^{n_{r-1}}\alpha^{''}a_j\le \sum_{j=1}^{n_{r-1}}b_j=\sum_{j=1}^{n_{r-1}}x_j$.
So
\begin{equation}
\sum_{j=n_{r-1}+1}^\eta \alpha^{''}a_j\ge \sum_{j=n_{r-1}+1}^\eta x_j=\sum_{n_{r-1}+1}^\eta \alpha_{n_{r-1}}a_j.
\end{equation}
Thus $\alpha^{''}\ge \alpha_{n_{r-1}}$, and hence
$\alpha^{''}a_\eta\ge \alpha_{n_{r-1}}a_{\eta}=x_\eta$.
Let $\hat{x}=(\alpha^{''}a_1,\ldots,\alpha^{''}a_\eta,x_{\eta+1},\ldots,x_n)$.
Then $\hat{x}$ is decreasing and $\hat{x}\prec b$.
By $(a)$, $n_1=\max\{k:(x_1,\ldots,x_k)=\alpha (a_1,\ldots,a_k)\mbox{ for some }\alpha\}$ and $n_1<\eta$.
Hence, $(x_1,\ldots,x_\eta)\neq \alpha^{''}(a_1,\ldots,a_\eta)$.
We also have $\alpha^{''}=\frac{x_1+\cdots+x_\eta}{a_1+\cdots+a_\eta}$.
By Lemma \ref{lemma}(b),
\begin{equation}
f(a)g(\hat{x})-f(a)g(x)=\sum_{j=1}^\eta f(a_j)g(\alpha^{''}a_j)-\sum_{j=1}^\eta f(a_j)g(x_j)>0.
\end{equation}
This is a contradiction.
Hence $n_1=\eta$.

\medskip
By induction, we only need to show the case $n_2$.
From the $n_1$ case, we have $\sum_{j=1}^{n_1}x_j=\sum_{j=1}^{n_1}b_j$.
Thus, $(x_{n_1+1},\ldots,x_n)\prec (b_{n_1+1},\ldots,b_n)$,
and
\begin{equation}
\sum_{j=n_1+1}^nf(a_j)g(x_j)\le \max_{(y_{n_1+1},\ldots,y_n)\prec
(b_{n_1+1},\ldots,b_n)}\sum_{j=n_1+1}^n f(a_j)g(y_j).
\end{equation}
On the other hand, for any
$(y_{n_1+1},\ldots,y_n)\prec (b_{n_1+1},\ldots,b_n),$ we have
$(x_1,\ldots,x_{n_1},y_{n_1+1},\ldots,y_n)\prec b$.
Then
\begin{eqnarray}
\sum_{j=1}^n f(a_j)g(x_j)&=&\max_{y\prec b}f(a)g(y)\nonumber\\
&\ge& \sum_{j=1}^{n_1}f(a_j)g(x_j)+ \max_{(y_{n_1+1},\ldots,y_n)\prec
(b_{n_1+1}, \ldots,b_n)}\sum_{j=n_1+1}^n f(a_j)g(y_j).
\end{eqnarray}
This implies that
\begin{equation}
\sum_{j=n_1+1}^nf(a_j)g(x_j)= \max_{(y_{n_1+1},\ldots,y_n)\prec
(b_{n_1+1},\ldots,b_n)}\sum_{j=n_1+1}^n f(a_j)g(y_j).
\end{equation}
From the proof of the case $n_1$, the result follows.
\qed

{\bf Proof of Theorem \ref{3.3}}
From Theorem \ref{3.2}, we need only to determine the maximum of $\sum_{j=1}^nf(a_j)g(x_j)$ for $x_1\ge \cdots\ge x_n\ge 0$ and $(x_1,\ldots,x_n)\prec (b_1,\ldots,b_n)$.
Suppose that $a_r>0$ and $a_{r+1}=\cdots=a_n=0$.
Let $\alpha\equiv \sum_{j=1}^nf(a_j)g(d_j)$ attain the maximum for $d_1\ge \cdots\ge d_n\ge 0$ and $ (d_1,\ldots,d_n)\prec (b_1,\ldots,b_n)$.
Then $\alpha=\sum_{j=1}^rf(a_j)g(d_j)$ and $(d_1,\ldots,d_r)\prec_w (b_1,\ldots,b_r)$.
Since $f$ is nonnegative and $g$ is increasing,
\begin{eqnarray}
&\max\{\sum_{j=1}^rf(a_j)g(x_j):x_1\ge \cdots\ge x_r\ge 0,(x_1,\ldots,x_r)\prec_w (b_1,\ldots,b_r)\}\\
&\le \max\{\sum_{j=1}^rf(a_j)g(x_j):x_1\ge \cdots\ge x_r\ge 0,(x_1,\ldots,x_r)\prec (b_1,\ldots,b_r)\}\equiv \beta.
\end{eqnarray}
Hence $\alpha\le \beta$.
Given $x_1\ge \cdots\ge x_r\ge 0$ and $(x_1,\ldots,x_r)\prec (b_1,\ldots,b_r)$, choose $(y_1,\ldots,y_n)=(x_1,\ldots,x_r,b_{r+1},\ldots,b_n)$.
Then $y_1\ge \cdots\ge y_n\ge 0$, $(y_1,\ldots,y_n)\prec (b_1,\ldots,b_n)$, and $\sum_{j=1}^rf(a_j)g(x_j)=\sum_{j=1}^nf(a_j)g(x_j)$.
We obtain $\alpha=\beta$.
By Theorem \ref{concave}, we see that the algorithm will produce
the state of the form $\Phi(\rho_2)$ attaining the maximum. \qed

\section{Concluding remarks and further research}

Let $(\sigma_1,\sigma_2) \mapsto D(\sigma_1, \sigma_2)$ be a scalar function on quantum states $\rho_1, \rho_2$, such as
the trace distance, the fidelity function, and the relative entropy.
 For two given quantum states $\rho_1,\rho_2$, we determine optimal bounds for
$D(\rho_1, \Phi(\rho_2))$ for  $\Phi \in \mathcal{S}$
for different classes of functions $D(\cdot, \cdot)$,
where $\mathcal{S}$ is the set of unitary quantum channels, the set of mixed unitary
channels, the set of unital quantum channels, and the set of all quantum channels.
Specifically, we obtain results for functions of the following form

\begin{itemize}
\item[(a)] $D(\sigma_1,\sigma_2) = d(\lambda(\sigma_1-\sigma_2)),$
where $d(X)$ is a Schur-convex function on the eigenvalues of $X \in \mathcal{H}_n$,
\item[(b)] $D(\sigma_1,\sigma_2) = {\rm tr}(f(\sigma_1)g(\sigma_2))$, and
$D(\sigma_1,\sigma_2) = {\rm tr}|f(\sigma_1)g(\sigma_2)|$, where
$f,g: [0,1]\rightarrow  \mathbf{R}$.
\end{itemize}

For the class of function in (a), optimal bounds for $D(\rho_1, \Phi(\rho_2))$
are given for $\Phi \in \mathcal{S}$ for the four classes of quantum channels mentioned above.
Actually, the results and techniques in Section 2
can be extended to functions of the form
\begin{equation}
D(\sigma_1,\sigma_2) = d(\lambda(\alpha \sigma_1 - \beta \sigma_2))
\end{equation}
for given $\alpha, \beta \in \mathbf{R}$, and a Schur-convex function
$d$.

For the class of functions in (b), the optimal lower and upper bounds for
 $D(\rho_1, \Phi(\rho_2))$  are given for $\Phi\in \mathcal{S}$, where $\mathcal{S}$ is
 the set of unitary channels.   For the set of mixed unitary channels,
 the set of unital channels, and the set of all quantum channels, we determine the
best lower bound if $g$ is an increasing concave function; we also  find the best
upper bounds for special functions including the fidelity and relative entropy functions.
The results and techniques in Section 3 can be extended
to cover functions $D: \mathcal{D}_n\times \mathcal{D}_n \rightarrow \mathbf{R}$  of the form
$D(\sigma_1,\sigma_2) = \psi(f(\sigma_1)g(\sigma_2))$, where
$\psi(X)$ is a Schur concave function on the singular
values (eigenvalues or diagonal entries) of the matrix $X$.

There are many related problems deserving further study.
For instance, one may consider Problem 1.1 for a wider class of functions $D$ and
different classes of $\mathcal{S}$. More generally,
one may study the optimal bounds for the set
\begin{equation}
\{D(\rho_1, \Phi(\sigma)): \Phi \in \mathcal{S}, \sigma \in \mathcal{T}\}
\end{equation}
for a set $\mathcal{S}$ of quantum channels, and a set $\mathcal{T}$ of quantum states.
If $\mathcal{T} = \{\sigma_1, \dots, \sigma_k\}$
is a finite set, then one can apply our results to
$D(\rho_1, \Phi(\sigma_j))$ for each $j$ to get the optimal bounds for each $j$,
and compare them.

\bigskip\noindent
{\large \bf Acknowledgment}

The authors would like to thank Professor Jianxin Chen for some helpful discussion.
Li is an affiliate member of the Institute for Quantum Computing, University of Waterloo.
He is also an honorary professor of the University of Hong Kong, and the
University of Shanghai. The research of Li was
supported by USA NSF grant DMS 1331021, Simons Foundation Grant 351047, and
and NNSF of China Grant 11571220. The research
of Wang was supported by the Ministry of Science and Technology of the Republic of China under
project MOST 104-2918-I-009-001

\end{document}